\documentstyle[prl,aps,twocolumn]{revtex}
\newcommand{\beq}{\begin{equation}}
\newcommand{\eeq}{\end{equation}}

\def\eqa{\begin{eqnarray}}
\def\eea{\end{eqnarray}}
\begin{document}
\draft
\flushbottom
\twocolumn[
\hsize\textwidth\columnwidth\hsize\csname @twocolumnfalse\endcsname
\title{ Hamiltonian Description of Composite Fermions: Aftermath }
\author{    R.Shankar  }
\address{
 Department of Physics, Yale
University, New Haven CT 06520}
\date{\today}
\maketitle
%\tightenlines
%\widetext
%\advance\leftskip by 57pt
%\advance\rightskip by 57pt

\begin{abstract}

The  Lowest Landau Level (LLL), long distance  theory of Composite  
Fermions
(CF) developed by Murthy and myself is minimally extended to all  
distances,
guided by  very general principles. The resulting theory is   
mathematically
consistent, and physically  appealing: we  clearly see the electron  
and the
vortices binding to form the CF. The meaning of the   constraints,  
their role
in ensuring  compressibility of dipolar objects at $\nu =1/2$ , and   
the
observability of dipoles are clarified.

\end{abstract}
\vskip 1cm

]

%\narrowtext
%\tightenlines

Composite bosons (CB )\cite{CB}  and Composite fermions (CF) are  
dressed
versions of electrons that appear in the  LLL, long wavelength  
description of
the Fractional Quantum Hall States. CF  were introduced by  
Jain\cite{jain}  as
an organizing principle and a source of excellent trial wavefunctions  
for the
series $\nu = p/(2ps+1)$ to which we limit our attention.  The study  
of trial
wavefunctions  and   gedanken experiments  have yielded many CF  
properties such
as their charge $e^*$, dipole moment $d^*$, and magnetic moment  
$\mu^*$. This
paper is concerned with deriving these (and more)  from the  
primordial
electronic hamiltonian.

We begin by defining the CF. Laughlin\cite{laughlin} showed that  the  
quantum
Hall  ground states support excitations  called {\em vortices}. A   
vortex at
$z_0$ corresponds to an additional  factor $\prod_j (z_j-z_0)$ in the
wavefunction and  represents     a charge {\em deficit} of  
$p/(2ps+1)$ in
electronic units. {\em The CF is  $2s$ such vortices bound to an  
electron.} An
electron bound to $2s$ {\em flux tubes}, each of which corresponds to    
a
factor $\prod_j (z_j-z_0)/|(z_j-z_0) |$, will be called a  
Chern-Simons (CS)
fermion. Since a flux tube is neither charged nor a physical  
excitations,  the
CS fermions is not very physical.  (  Both Halperin
\cite{halperin1}  and
Read\cite{read1}  have emphasized the binding of electrons   to zeros  
or
vortices.)

The underlying   theory should somehow   introduce vortices and  
naturally bind
them to
electrons,  reproducing the   CF of charge $e^* = 1 -  2ps/(2ps+1) =
1/(2ps+1)$,
 dipole moment $d^* = -l^2 \hat{z}\times {\bf p}$ for $\nu =1/2$  
predicted by
Read\cite{read1} (and its  natural extension $\nu \ne 1/2$),
      magnetic moment $\mu^* = e/2m$,   predicted by Simon, Stern and
Halperin\cite{SSH} and   a    $1/m^*$ determined by interactions.

Chern-Simons (CS) theories in some  guise  are used to  relate     
electrons
to  the CF.
First one  makes the  Leinaas  and
Myrheim,\cite{LM} transformation  from the electronic wavefunction  
$\Psi_e$ to
$\Psi_ {CS}$, describing the CS fermions:

\begin{eqnarray}
\Psi_e &=&
\prod_{i<j} {(z_i - z_j) ^{2s}\over |z_i
-z_j|^{2s}}\Psi_{\rm CS}.\label{eq-phase}\\
H_{CS} &=& \sum_i {({\bf \Pi}_{i} + :{\bf a}_{cs}:)^2 \over 2m} +  V   
\\
{\bf \Pi }&=& {\bf p + eA^* }\ \ \ \     {\bf A}^* = {\bf  
A}/(2ps+1)\\
\nabla \times :{{\bf a}_{cs}}: &=& 2\pi s :\rho :
\end{eqnarray}
  where   $:{\bf a}_{cs}:$ and   $  :\rho :$ refer to fluctuation.

  Lopez and Fradkin \cite{frad}  to described
the Jain series  $\nu = p/(2ps+1)$ this way.    Kalmeyer and
Zhang\cite{kalmeyer} and in greater  depth,  Halperin, Lee and Read  
(HLR)
\cite{HLR} used it to make dramatic predictions  for  $\nu =1/2$,   
verified
experimentally\cite{willett}. The CS transformation  attaches
flux  and not vortices:\cite{RS} the CS fermion sees a weaker  
magnetic field
${\bf A}^*$, (and has a larger magnetic length $l^*$) but couples to  
scalar
potentials with unit charge and not $e^*$. These and  other problems   
listed in
Ref. \cite{Olle}
 led   Murthy and myself to propose  two further canonical  
transformations.  In
the first, collective coordinates $(A, A^{\dag})$ corresponding to  
gauge
fluctuations were introduced similar to   Bohm-Pines {\cite{BP} and  
found to
lead to oscillators   at the cyclotron  scale $\omega_0 = eB/m$. In  
this larger
Hilbert space, physical states were required to obey  constraints,  
one for each
oscillator. The second transformation decoupled the oscillators from  
the
fermions  in the infrared. This low energy, long-wavelength  theory,   
allowed
one to derive and reconcile  many  properties of the CF. However it  
too had
some troubling  questions  to be addressed here.

 I propose  an extension of that theory to all length scales based on  
very
general  principles.  The result is mathematically sound and very     
obviously
describes    CF's.

After  our  last  transformation we found at small $ql$
 \begin{eqnarray}
H &=& \sum_{q}^{Q}\omega_{\rm 0}
A^{\dag}(q)A(q)  +\sum_j {eB^*\over 2m}\nonumber \\
 \!\!\!\! &+&  \left[  \sum_j {\Pi_{-}^{j}\Pi_{+}^{j}\over
2m} \!
 - \!\! {1 \over 2mn}\! \sum_i\sum_j
\sum_{q}^{Q}
\Pi_{-}^{i}e^{-i{\bf q \cdot (x_i-x_j)}} \Pi_{+}^{j}\right] \!\! \! +  
\!\!  V
\nonumber \\
&\equiv & H_{osc} + H_{\mu} + H_{0}  + V\label{H}
\end{eqnarray}
where $\Pi_{\pm} = \Pi_x \pm i \Pi_y$.  For LLL physics we  
ignored\cite{drop}
$H_{osc}$  (the  oscillators were  assumed  frozen in the ground  
state with
$\langle A\rangle=\langle A^{\dag}\rangle=0$), and   dropped the   
magnetic
moment term $H_{\mu}$. As for
$H_{0}$, since  the $i=j$ terms from the second sum  renormalize the  
mass as
follows:
\beq
{1 \over m^*} = {1 \over m} \left( 1- {1\over n}\sum_{q}^{Q}\right)
\label{massren}
\eeq
the kinetic energy is exactly quenched if we choose $Q =k_f$. The  
$i\ne j$
terms were be  shown to be convertible to  an interaction with  
support at
$q>Q$, which, when combined with the ${\bf :a}_{CS}:$ which exists  
only for
these values, produces  the plasmon pole with correct location and  
residue. In
the infrared, the hamiltonian, electron density $\rho^e$,  and  
constraint
$\bar{\chi}$, became
\begin{eqnarray}
H &=&  V  = {1 \over 2} \sum_{\bf q} \rho ({\bf q}) v(q) \rho (-{\bf  
q})\\
{\rho}^e (\bf{ q})  &=& \underbrace{ \sum_j e^{-i{\bf q \cdot
r}_j}\left[ 1 -  {i\l_{}^{2}\over (1+c) } {\bf q}\times
{\bf \Pi}_{j} \right]}_{ {\mbox{\Large $\bar{\rho}$}}} + {Osc}  
\label{rho}
\nonumber \\
\bar{\chi}(\bf{q})  &=&  \sum_j  e^{-i{\bf q \cdot
r}_j}\left[  1 +  {i\l_{}^{2}\over c(1+c) }
{\bf q}\times {\bf \Pi}_{j} \right]= 0 \ \   \label{series}\\
c^2 &=& {2ps\over 2ps+1}=2\nu s
\end{eqnarray}
where $Osc$ refers to an  $A + A^{\dag}$ term, to be dropped. Thus
$\bar{\rho}$  is the LLL projected electron density.\cite{drop}

Let us begin with the positive aspects of these results.  They were  
{\em
derived}  from the electronic hamiltonian. {\em    They hold for  all  
Jain
fractions}. The constraints do not involve the oscillators.   They  
were derived
independently by D. H. Lee,
 (at small $ql$) starting with bosons.\cite{DH}
 Finally, Pasquier and
Haldane\cite{PH}  and Read\cite{read2}    corroborates some  of these
results as will be explained later.

Now for the problems.   Our $\bar{\rho}$ has transition matrix  
elements of
order $q$ rather than $q^2$ between  free particle  CF states,   
violating
Kohn's theorem.  Constraints must  somehow be  incorporated  to
fix this. Next, we  know from Girvin, MacDonald and Platzman  
(GMP)\cite{GMP}
that the LLL
projected charge  should obey\cite{proj}
\beq \left[ \bar{\rho}({\bf q}) , \bar{\rho}({\bf q}') \right] =  \  
2i \sin
\left[ {l_{}^{2} ({\bf q\times q'}) \over 2}\right] \bar{\rho} ({\bf
q+q'}).  \label{GMP}
 \eeq
This algebra has a natural small $ql$ limit GMP2
\beq
\left[ \bar{\rho}({\bf q}) , \bar{\rho}({\bf q}') \right] = i
l_{}^{2} ({\bf q\times q'}) \bar{\rho} ({\bf q+q}')    \
\mbox{(GMP2)}.\label{GMP2}
 \eeq
Unfortunately  we find
\beq
\left[ \bar{\rho}({\bf q}) , \bar{\rho}({\bf q}') \right] \ne  i
l_{}^{2} {\bf (q \times q'})\bar{\rho}({\bf (q + q}) .
\eeq
Although the structure constant agrees with GMP2,  the  term of order
$qq'(q+q')$ does not correspond to
$\bar{\rho}({\bf q+q'})$  due to neglected  higher order terms.

We similarly find that  $\left[ \bar{{\chi}}({\bf q}) ,   
\bar{\rho}({\bf q}')
\right] \ne 0$ except to leading order, which means the charge is not  
gauge
invariant. It is not even  {\em weakly gauge invariant}, i.e.,  obey
$\left[ \bar{{\chi}} ,  \bar{\rho}\right] \simeq \bar{\chi}$ which  
would
ensure that $\bar{\rho}$ and $H(\bar{\rho})$ do not mix physical and  
unphysical
states. But there is a deeper problem:
 \beq
\left[ \bar{{\chi}}({\bf q}) ,  \bar{\chi}({\bf q}') \right] \ne
\mbox{(structure constant)} \bar{{\chi}}({\bf q+q'}),
\eeq
  because    ${\bf (q +q') \times \Pi} $ has the wrong coefficient    
to yield
$\bar{\chi} ({\bf q+q'})$.   Now,    $\bar{\chi}$ must close under  
commutation:
one cannot have
 individual factors in the commutator annihilate physical states but  
not their
commutator!   While    gauge invariance can be implemented order by  
order in  a
coupling
constant , this is not so with respect to   $q$ which is integrated  
over. (One
{\em can}  use   $Q$ as a
small parameter, though some   of the physics   gets  
modified.\cite{HS1,simon})

{\em I will now propose the  minimal    extension of our results  to  
all $ql$
that  is mathematically and physically attractive. }

Let us assume that    Eqn.(\ref{series})  represents  the beginnings  
of   two
exponential series and adopt  the following  expressions for charge  
and
constraint:
\begin{eqnarray}
\bar{\bar{\rho}} &=& \! \sum_j \exp (-i{\bf  q} \! \cdot \! ({\bf   
r}_j\! - \!
{l^2\over 1+c}\hat{\bf z}\times {\bf \Pi}_j ))\equiv \sum_je^{-i{\bf  
q\cdot
R_{ej}}}\label{robar}\\
\bar{\bar{\chi}} &=& \! \sum_j \exp (-i{\bf  q}\! \cdot  \!({\bf   
r}_j \!+
\!{l^2\over c(1+c)}\hat{\bf z}\times {\bf \Pi}_j ))\equiv \! \sum_j  
e^{-i{\bf
q\cdot R_{vj}}}\label{chibar}
\end{eqnarray}
{\em Note that  ${\bf R}_e\ \mbox{and}\  {\bf R}_{v} $ were fully  
determined by
the two terms we did  derive}.

Now to reap the benefits. First note that  given
\begin{eqnarray}
{\bf R}_e &=& {\bf r} -{l^2\over (1+c)}\hat{\bf z}\times {\bf \Pi  
}\label{re},
\\
\left[ R_{ex}\ , R_{ey} \right] &=& - il^2.
\end{eqnarray}

These describe the guiding center coordinates of a unit charge  
object, which
is clearly the electron. (Note also     that  ${\bf R}_e$ enters the  
formula
for projected electron density.)   We can
readily   combine  exponentials  and   show that     
$\bar{\bar{\rho}}({\bf q})$
 obeys  the   GMP algebra, Eqn.(\ref{GMP}).

Next consider the other set of coordinates
\begin{eqnarray}
{\bf R}_v &=& {\bf r} +{l^2\over c(1+c)}\hat{\bf z}\times {\bf \Pi
}\label{rv}\\
\left[ R_{vx}\ , R_{vy} \right] &=&  il^2/c^2.
\end{eqnarray}
These   describe the guiding center coordinates of a particle whose
charge is $-c^2$, namely the $2s$-fold vortex. It follows that  
constraints
$\bar{\bar{\chi}}(q)$ also close to form a GMP algebra with $l^{2}\to   
-
l^2/c^2$ in the  structure constants.

Finally, and fortunately,  $\left[ \bar{\bar{\chi}}\ ,  
\bar{\bar{\rho}} \right]
=0 =\left[ \bar{\bar{\chi}}\ , H(\bar{\bar{\rho}}) \right] $   since
\beq
\left[ {\bf R}_e\ , {\bf R}_{v} \right] = 0. \label{evcomm}
\eeq
  As in the  Yang-Mills case,  the constraints form a
nonabelian algebra and commute with $H$.

I do not imply that an  exact implementation of our last canonical
transformation will lead to the above results; I know this is  not  
so.  I have
minimally extended the (derivable)
small $ql$ theory to all $ql$,  guided by mathematical      
consistency.
The resulting short distance physics
could well  be  at odds with CF maxims, but isn't.

Consider   Eqn. (\ref{re}, \ref{rv}). They show us the innards of the  
CF rather
explicitly: a CF at ${\bf r}$ with kinetic momentum ${\bf \Pi}$,  is  
flanked by
the electron and vortex within a distance of order $l^2 \Pi$.  Its  
total charge
is  their sum $e^* = 1/(2ps+1)$. Its dipole moment (in the frame  
${\bf r}=0$)
is  $d^* = -l^2 \hat{\bf z}\times {\bf \Pi}$. Its size $l^2 \Pi
\simeq l$  near the Fermi surface,  making it a well  defined object  
in this
energy range.  (These discussions involving operators are   
semiclassical. In
addition, ${\bf \Pi}$ is not a constant of motion except at $\nu  
=1/2$ when it
equals ${\bf p}$.) The four dimensional phase
space of the CF  has spawned  two guiding center coordinates: ${\bf  
R}_e$ and
${\bf
R}_v$ which  is twice as many coordinates for the {\em electronic LLL
problem. }  {\em But the constraints  tell us the density formed out  
of the
vortex coordinates has no fluctuations, analogous to the Bohm-Pines  
condition
$\sum_j \exp {i{\bf q \cdot r}_j} = 0 $ for the small $q's$  at   
which plasmons
are introduced}.  The LLL condition on electrons is imposed by  
freezing the
oscillators  in their ground states.

Halperin and Stern have always maintained  that   dipolar fermions  
are
  compressible,  just like the unit charge CS  fermions in  HLR. The  
proof was
sketched  in \cite{HS1} for $H_0$ of Eqn.(\ref{H}) and in generality  
by Stern
{\em
et al}\cite{simon}. D.H.Lee came to the same conclusion in  
Ref.\cite{DH}.
Finite compressibility was   established for the related $\nu =1$  
bosons
problem by Read\cite{read2}.    All these proofs relied on a careful
implementation
of the constraints or     gauge invariance.  We may now understand  
this  as a
follows:    {\em $\bar{\bar{\chi}}=0$  means  that the vortex  ends  
of the
dipoles have  no
collective density fluctuations.}   Thus  only the electron ends  
respond  to
the  static potential, exhibiting  static  compressibility.

{\em This does not however mean the dipole is a red herring.   As we  
move up in
frequency, the dipoles begin to act independently of each other and  
of the
constraint.} I base this on  an analysis of   Read's work  
\cite{read2} on
the  irreducible density-density response function $K^{\rm irr}$ for  
$\nu=1$
bosons,  relevant to us because  their  LLL description {\em a la}     
Pasquier
and Haldane\cite{PH},  is a fermionic theory in which the charge and  
constraint
coincide with  Eqns. (\ref{robar}-\ref{chibar})   upon setting  $c=1,  
{\bf \Pi
=p}$, which corresponds to $\nu =1/2$.  Any response function  
computed for one
problem is readily  transformed to the other upon taking  into  
account the
trivial difference in  magnetic lengths. Thus one  begins with
\begin{eqnarray}
H &=& {1 \over 2} \sum \bar{\bar{\rho}}\  v(q)e^{-(ql)^2/2}\   
\bar{\bar{\rho}}
  \ \ \ \  \left[ H , \bar{\bar{\chi}} \right] =0  \ \ \   
\bar{\bar{\chi}} =0
\label{natural}
\end{eqnarray}
where $e^{-(ql)^2/2}$ takes into account the fact that  
$\bar{\bar{\rho}}$ is
really the magnetic translation and not projected density, a  
difference that
did  not matter in the earlier small $ql$ work\cite{Olle}.
  In the conserving approximation, the inverse  mass is zero at tree  
level and
arises
from the Fock diagram. (If $H$ is expanded in first quantization  
there is no
$p^2/2m$ term at $\nu =1/2$).  The vertex is dressed by the  
corresponding
ladder. A series expansion of $\bar{\bar{\rho}}$ (see Eqn.  
(\ref{rho}))  shows
unit charge and half the final  dipole moment. Implementing  the  
constraint in
a conserving approximation leads to a gauge field whose longitudinal  
part $a_L$
screens the charge fully,  leaving behind the right dipoles which  
then interact
via the transverse gauge field $a_T$.   The result is
\beq
K^{\rm irr} = K^{\rm irr}_{dipole} + K^{\rm irr}_{\rm tgf}\label{K}
\eeq
The first  piece (at small $ql$) is  simply the dipole-dipole  
correlation of a
 Fermi sea. In the second term   these dipoles couple
via  a  transverse gauge field.  As $\omega \to 0$, $K^{\rm irr}_{\rm
tgf}$   dominates, due to the singular propagator of $a_T$  giving  
nonzero
compressibility. (The  field $a_T$ also produces mass divergences at  
the Fermi
energy as in HLR).
As $\omega$ increases,   the the situation changes: at
$\omega \simeq  qv_F$, the first term is twice as big and eventually
dominates. In LLL sum rules,  the $\omega^{-1} $ moment is  
incorrectly given by
 free dipoles, while the zeroth moment agrees  (up to logarithms),  
and   the
positive moment is saturated by them.   {\em Thus, away from the low   
frequency
region, the  answer is given by the  correlation function of   
non-interacting
dipoles.}  For  gapped fractions (not too close to $\nu =1/2$) a  
description in
terms of  independent particle with  $e^*$ and $d^*$  is likewise  
expected to
be a good approximation, perhaps for all $\omega$,     since the gap  
will
cut-off the low frequency end (so that  $a_T$ can't raise its head)  
while  the
major effects of $a_L$ are already encoded in $e^*$ and $d^*$.

  Rather than reach this description  using the conserving  
approximation (which
could be very difficult in the presence of LL structure) I  propose   
a scheme
in which $e^*$ and $d^*$ are built in at tree and so that  additional  
effects
of constraints should be very small. The scheme  is  inspired by our  
work
\cite{Olle}):   use   the {\em preferred combination} for charge  
density
\beq
\bar{\bar{\rho}}^p = \bar{\bar{\rho }} - c^2 \bar{\bar{\chi}}
\eeq
physically  equivalent to $\bar{\bar{\rho }}$ and {\em weakly gauge  
invariant}:

\beq
\left[ \bar{\bar{\chi}} \ , \bar{\bar{\rho}}^p\right] \simeq  
\bar{\bar{\chi}}.
\eeq
Clearly so is the  $H(\bar{\bar{\rho}}^p) $\ that I begin with:
\begin{eqnarray}
H^p &=&  {1 \over 2} \sum \bar{\bar{\rho}}^p \ v(q)e^{-(ql)^2/2} \
\bar{\bar{\rho}}^p ;   \ \   \left[ H^p , \bar{\bar{\chi}} \right]  
\simeq
\bar{\bar{\chi}}  \ \ \  \bar{\bar{\chi}} =0   \label{preferred}
\end{eqnarray}

 Consider the series expansion of  $\bar{\bar{\rho}}^p$:
\beq
\bar{\bar{\rho}}^p = \sum_j e^{-i{\bf q \cdot r}_j}\!\! \left( \! {1  
\over
2ps+1}\!  -  {i\l_{}^{2}  }
q\times
{\bf \Pi}_{j}  \! +{0} \cdot
\left( q\times
{\bf \Pi}_{j} \right)^2 \! + \cdots \right) \label{rostar}
\eeq
Note that  $\bar{\bar{\rho}}^p$  has     $e^*$ and $d^*$  (and hence
$q^2$ matrix elements)\cite{GM}  built in and that $H^p$ contains a  
$p^2/2m^*$
term for each particle at $\nu =1/2$  (and a $\Pi^2/2m^*$ at other  
fractions)
where  $1/m^*$ is determined by the interactions.

As an illustration consider our  computation of  gaps\cite{gaps}. We   
simply
sandwiched $H^p$ between the CF ground state ($p$ filled LL)  and a
particle-hole excitation of it, and took the difference. In this   
small $ql$
treatment, we used
\beq
\bar{\rho}^p = \sum_j e^{-i{\bf q \cdot r}_j}\!\! \left( \! {1 \over  
2ps+1}\!
-  {i\l_{}^{2}  }
q\times
{\bf \Pi}_{j}  \right)
\eeq
and not the full series $\bar{\bar{\rho}}$, which was not in the  
picture then.
We had chosen  this combination of $\bar{\rho}$ and $\bar{\chi}$ from  
Eqn.
(\ref{rho})  since it had    the   correct $e^*$ and $d^*$  (and  
hence
$q^2$ matrix elements)\cite{GM}  built in and also obeyed GMP2.   Our   
scaling
laws relating  gaps
at the same $p$ but different $s$ agreed very well with numerical  
work .

 We now understand why $\bar{\rho}^p$  worked so well:
 $\bar{\bar{\rho}}^p =\bar{\bar{\rho }} - c^2 \bar{\bar{\chi}}
 $, captures so much of CF physics because,  {\em  being  the sum of
electron
and vortex densities weighted by their charges, it  is   the CF  
charge
density!}
That  $\bar{\bar{\rho}}^p$ has a vanishing third
moment and  the fourth   is down by a factor of at least 500
 near  the Fermi surface,  explain why   $\bar{\rho}^p$ obeys GMP2  
and works
over a wide  range of $ql$.

   While scaling laws for gap ratios worked very well, the gaps  
themselves
diverged for coulomb interactions on samples of  thickness $\Lambda  
=0$ because
we could not consistently include the  $e^{-(ql)^2/2}$ factors in  
$H^p$ in a
small $q$ formalism. (Instead we used a LL cut-off on CF states.)  
However for
$\Lambda > 0$,  the factor $e^{-q\Lambda} $ that represented  
thickness in
$v(q)$   killed off large $ql$ and the numbers  agreed very well  
(within
$20\%$)  with Park and Jain  beyond $\Lambda \simeq 2 l$. I have  
since
analytically computed the gaps using $H^p(\bar{\bar{\rho}}^p) $, and  
obtained
finite results for all $\Lambda$  and agreement to within $20\%$ or  
better for
$\Lambda \ge l/2$ for $\nu =1/3,2/5$ and $ 3/7$. I  calculated  in  
closed form
the profiles of quasielectrons and holes simply by evaluating
$\bar{\bar{\rho}}^p$ in states  with an extra CF or hole on top of  
$p$ -filled
LL's. These agree with  Park and Jain (unpublished) in their  size  
and
oscillation scale  ( set by $l^*$),  but are off by $10-30\%$ in the
amplitudes.   As for  $S(q)$, the oscillations have the right  
wavelength but
very small amplitudes   compared to Monte-Carlo work\cite{kam}.

  I   have presented  a formulation applicable to   the entire Jain
series,  based  what we could  derive at small $ql$,  and   minimally   
extended
to all $ql$ using very general  consistency principles. The resulting  
theory
has  electrons bound to vortices by energetics to form CF of the  
right charge
and dipole moment. The role of the constraints and dipoles is  
clarified. I have
proposed
two formulations, in terms of $H(\bar{\bar{\rho}}) $ and
$H^p(\bar{\bar{\rho}}^p) $,
 equivalent in  exact calculations, but suited     for different
approximations. The latter provides an {\em analytic} (but  
approximate)  scheme
for computing   gaps,   particle-hole profiles and structure factors  
directly
in terms of CF (without  transforming back to electrons).

I  thank J. Jain,  G. Murthy, N. Read and A. Stern for discussions  
and
the  NSF for grant DMR98-00626.


\begin{thebibliography}{99}
\bibitem{proj} If in the electron density $e^{-i{\bf q \cdot r}} =  
e^{-i{\bf q
\cdot R}}\ e^{-i{\bf q \cdot \eta}}$ we
simply drop the cyclotron coordinate ${\bf \eta}$  we get the  
magnetic
translations, while if we set  $ e^{-i{\bf q \cdot \eta}} = \langle  
e^{-i{\bf q
\cdot \eta}} \rangle_0 = e^{-(ql)^2/4}$, we get the projected  
density. Since we
dropped the oscillators altogether we are working with magnetic  
translations
and  Eqn.(\ref{GMP}) is their algebra.
\bibitem{GM2} G. Murthy  points out that   if $\bar{\rho}^p$ has  
$q^2$ matrix
elements,    GMP2  must follow to avert a   mismatch in powers of  
$q$.
\end{thebibliography}
\end{document}